# ON THE ORIGIN OF THE DOMINANT WAVES IN THE EXTENDED SOLAR CORONA

F.S. Mozer[1], O.V. Agapitov[1], O. Romeo[1], V. Roytershteyn[2], A. Voshchepynets[3]
[1]Space Sciences Laboratory, University of California, Berkeley, 94720, USA
[2]Space Science Institute, Boulder, Colorado, 80301, USA
[3]Department of System Analysis and Optimization Theory, Uzhhorod National University, Uzhhorod, Ukraine

The extended solar corona at 10-30 solar radii is essentially devoid of all waves below 100 kHz other than triggered ion acoustic waves (TIAW), which consist of a low frequency electromagnetic wave at a frequency of a few Hz coupled to one or more electrostatic waves at a few hundred Hz, such that the amplitudes of the higher frequency waves peak at a fixed phase of each low frequency wave period. All the waves in a TIAW event travel at the same phase speed, which is found to be 150 km/sec (the ion acoustic speed was about 100 km/sec). It has not been possible to explain the TIAW as a resonant wave-wave interaction, so a non-resonant interaction has been considered in which the loss of energy by the low frequency wave is used to both heat the electrons and grow the higher frequency waves. Evidence in support of this explanation is described and a PIC simulation that discusses this process is summarized. This simulation demonstrates wave-particle interactions driven by the enhanced LF fluctuations, which subsequently modify the proton VDF and create conditions favorable for the growth of HF waves. This interplay between a pair of waves, mediated by modifications of plasma parameters and energy conversion, represents a significant nonlinear process in plasma physics, the study of which will deepen the understanding of energy transfer, wave generation, and plasma dynamics in diverse astrophysical environments.

## I INTRODUCTION

The most intense and important wave in the outer solar corona (10-30 solar radii) is the triggered ion acoustic wave [Mozer et al, 2021; 2022; 2023; 2025]. The generation and interaction of these TIAW waves has not been understood in terms of a resonant wave-wave interaction so non-resonant interactions have been considered. Resonant interactions require that the wave frequency, $\boldsymbol{\omega}$, and the wave vector, $\mathbf{k}$, satisfy $\boldsymbol{\omega}_0=\boldsymbol{\omega}_1+\boldsymbol{\omega}_2$ and $\mathbf{k}_0=\mathbf{k}_1+\mathbf{k}_2$, while, for non-resonant interactions such restrictions are eliminated. Because the stringent matching conditions of a resonant interaction are relaxed for non-resonant interactions, the parameter space for such interactions is vastly larger than for resonant interactions. Essentially, any combination of waves present in the plasma can potentially interact non-resonantly. For example, the low frequency wave of the pair may lose energy due to Landau damping and the released energy may be used to both heat the electrons and grow the higher frequency wave. In this way, the coupling between the two waves results from the modification of the plasma medium by one wave that affects the growth rate of the other wave. The study of such interactions is vital for a comprehensive understanding of how wave energy is distributed and dissipated in plasma environments.

TIAW properties are described in section II. A particle-in-cell simulation, illustrating one possible coupling mechanism, is summarized in section III and section IV discusses some conclusions of this work. The electric and magnetic field data [Bale, et al, 2016] and the plasma electron data [Whittlesey et al, 2020] were obtained on the Parker Solar Probe. Such data is presented in the spacecraft coordinates with the Z- axis radially towards the Sun.

## II PROPERTIES OF TRIGGERED ION ACOUSTIC WAVES

Before the launch of the Parker Solar Probe (PSP), it was widely believed that the inner heliosphere was populated by whistler waves that heated and accelerated the outflowing plasma. However, PSP found that there were few if any whistlers inside a distance of 30 solar radii from the Sun [Cattell et al, 2021]. Instead, this region is populated by triggered ion acoustic waves, an example of which is given in Figure 1, in which panels (1b) and (1c) present a segment of a low frequency (2-10 Hz) and higher frequency (>100 Hz) wave that are linked together such that the amplitude of the higher frequency wave peaks at each fixed phase of the low frequency wave. As shown in panel (1a), this occurs because the higher frequency wave consists of several narrow band components that are separated in frequency by the frequency of the low frequency wave.

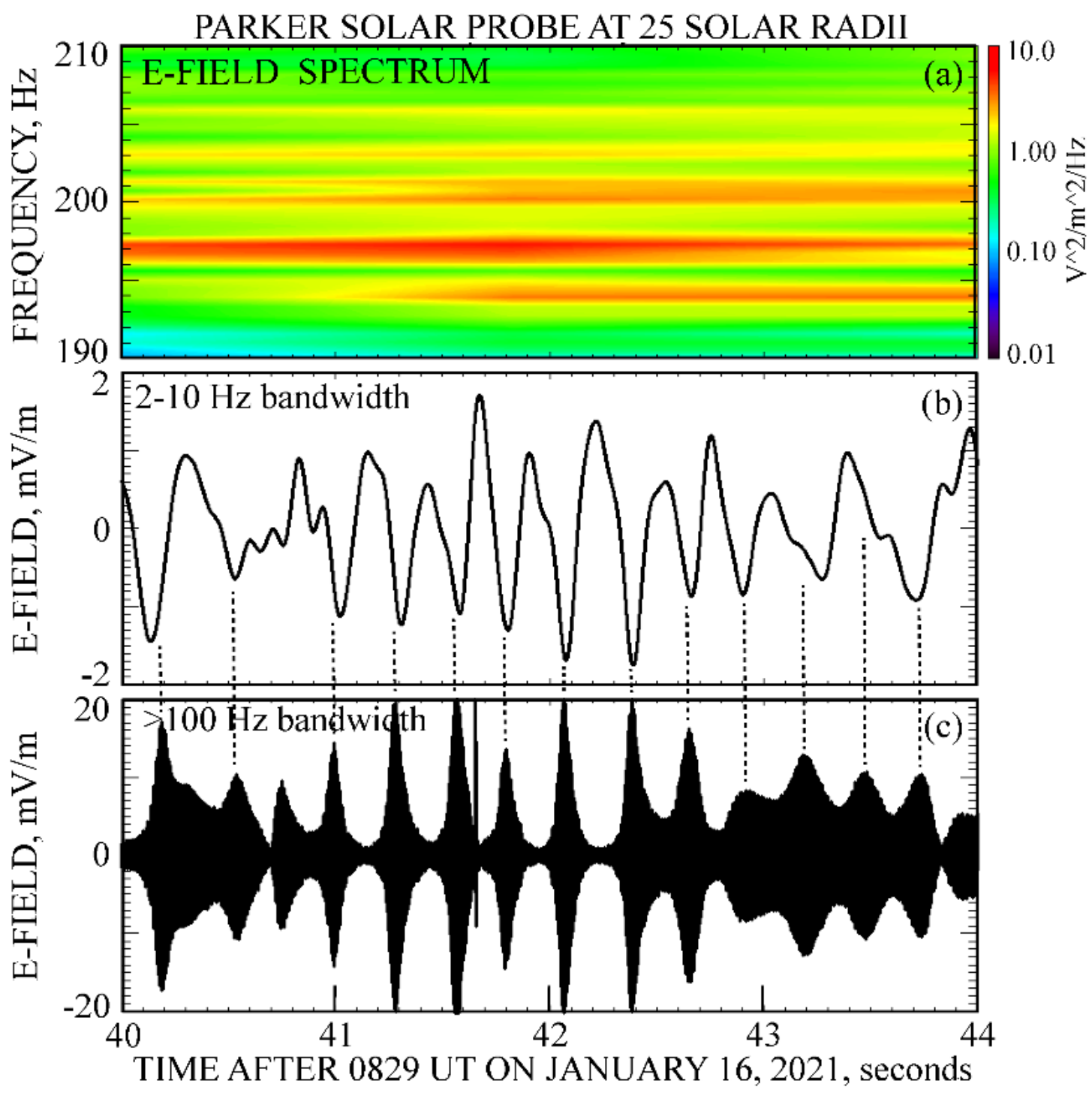

Figure 1. An example of triggered ion acoustic waves. Panel 1b gives the waveform of an ~4 Hz wave, while panel 1c illustrates the higher frequency wave that peaks at a fixed phase of the low frequency wave (see vertical dashed lines). Panel 1a, which provides the spectrum of the higher frequency wave, shows that this wave consists of several narrow band components that differ in frequency by the frequency of the low frequency wave.

Properties of these waves are:

1. **Coupled Wave Pair:** TIAW are characterized by the simultaneous presence of two distinct wave components: a low-frequency (LF) wave typically oscillating at a few Hz, and a high-frequency (HF) wave with frequencies of a few hundred Hz.
2. **Phase-Locked Coupling:** A crucial feature is the strong phase relationship between these two components. The HF wave often contains a short interval of larger amplitude that occurs during successive periods of the LF wave, and critically, at a fixed phase of this LF sinusoidal oscillation. This phase-locking is a primary indicator of a nonlinear coupling process rather than a coincidental superposition of independent waves.
3. **Narrow Bandwidth:** Both wave components exhibit remarkably narrow bandwidth components that appear as nearly pure sinusoidal waves
4. **Persistence:** TIAW occur between 18 and 30 solar radii during 75% of the Parker Solar Probe passes through this region on early Parker Solar Probe orbits and on all passes in later orbits, and, when present, they last for hours as the dominant wave mode.
5. **Association with Electron Heating:** In the presence of TIAW, electrons are heated while, in their absence, there is little or no electron heating between 18 and 30 solar radii [Mozer et al, 2022].

The narrow bandwidth and persistent phase-locking of TIAW strongly suggest a highly selective and sustained nonlinear triggered process, rather than stochastic wave generation from broad turbulence. Their effectively pure sine wave nature implies a coherent source or a strong filtering/resonance mechanism. Furthermore, phase-locking over extended periods, sometimes many hours, indicates that the conditions for this specific coupling are maintained or repeatedly met. This points away from random interactions and towards a deterministic nonlinear, non-resonant system where the LF wave acts as a precise modulator or trigger for the HF wave.

The low frequency wave of the TIAW is an electromagnetic wave, as shown in Figure 2 by the appearance of both a magnetic field component (panel 2e) and density fluctuations (panel 2d) in a narrow band electric field at 24 Hz (panels 2a, 2b, and 2c). As seen in panel (2g), the wave electric field is pointed at ~60° with respect to the background magnetic field (the red line) and the wave magnetic field is aligned at near 90° to the background magnetic field in panel (2f). Such waves have previously been reported in the solar wind [Jian et al, 2014; Ofman et al, 1999, 2012; Zieger et al, 2020]. The proton gyrofrequency during this event was equal to the 24 Hz frequency of this low frequency wave.

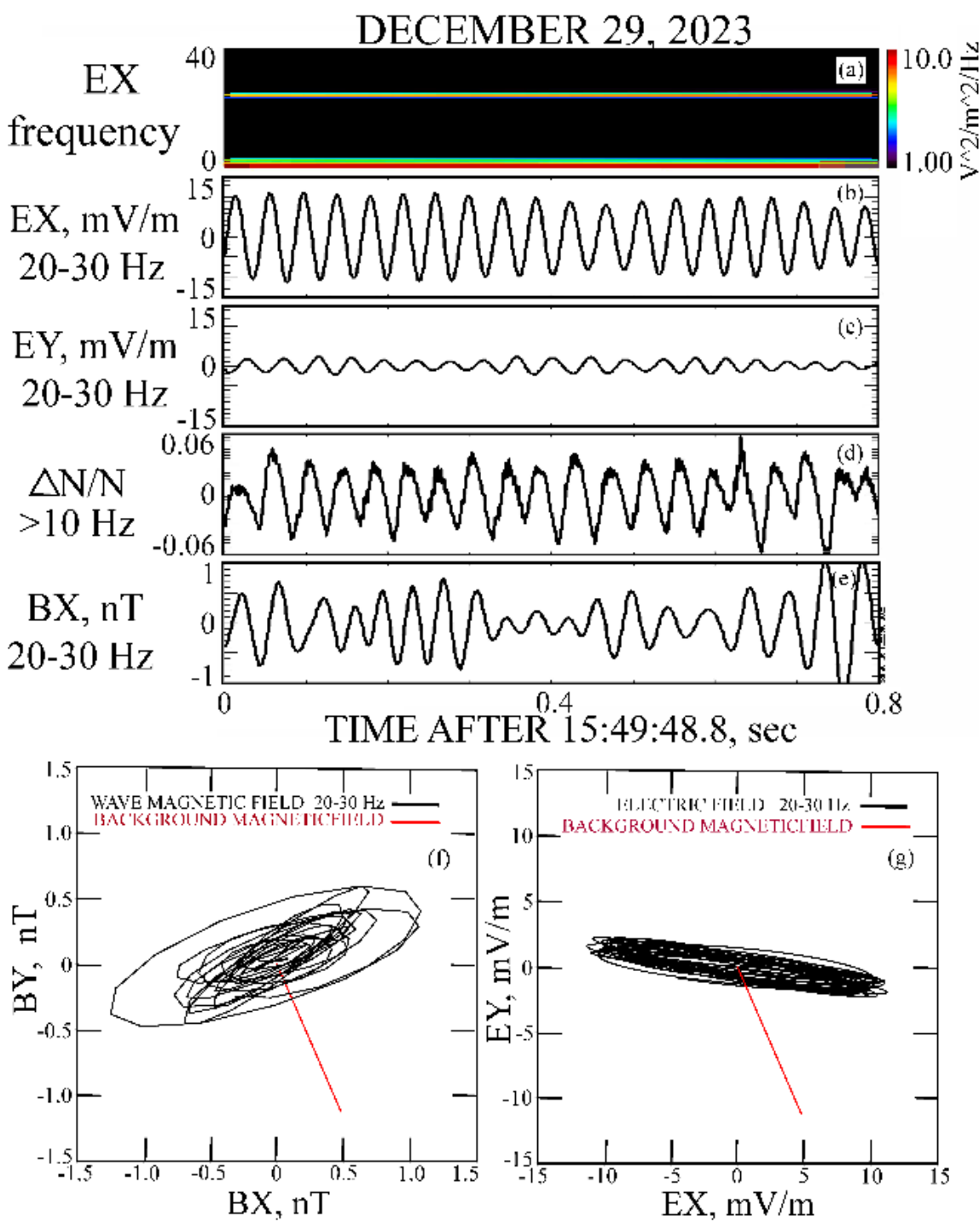

Figure 2. The low frequency wave of the TIAW is an electromagnetic wave because it contains both a magnetic field in panel (2e) and density fluctuations in panel (2d). Its electric field oscillates at approximately 60° with respect to the background magnetic field direction (panel 2g) while the wave magnetic field is at ~90° to the background B (panel 2f). The waveforms and spectra are illustrated in panels (2a), (2b), and (2c).

Attempts were made to further identify the low frequency wave mode by studying hodograms (curves or paths described by the tips of vectors) of the perpendicular magnetic and electric field components in the waves. While such plots might reveal correlations between pairs of functions, these studies revealed no well-defined and persistent correlations between the phases of the components and this was not due to a limitation of the measurements. Instead, it suggests that there is a mixture of left-hand and right-hand polarizations in the waveforms.

The high frequency component of the TIAW is an electrostatic wave because it contains density fluctuations (panel 3d) and no wave magnetic field (panel 3e). This wave consists of several narrow band signals that are separated in frequency by the low frequency wave frequency (panel 3a) and it propagates at about 10° relative to the background magnetic field (panel 3f). These properties suggest that the high frequency wave is an ion acoustic wave.

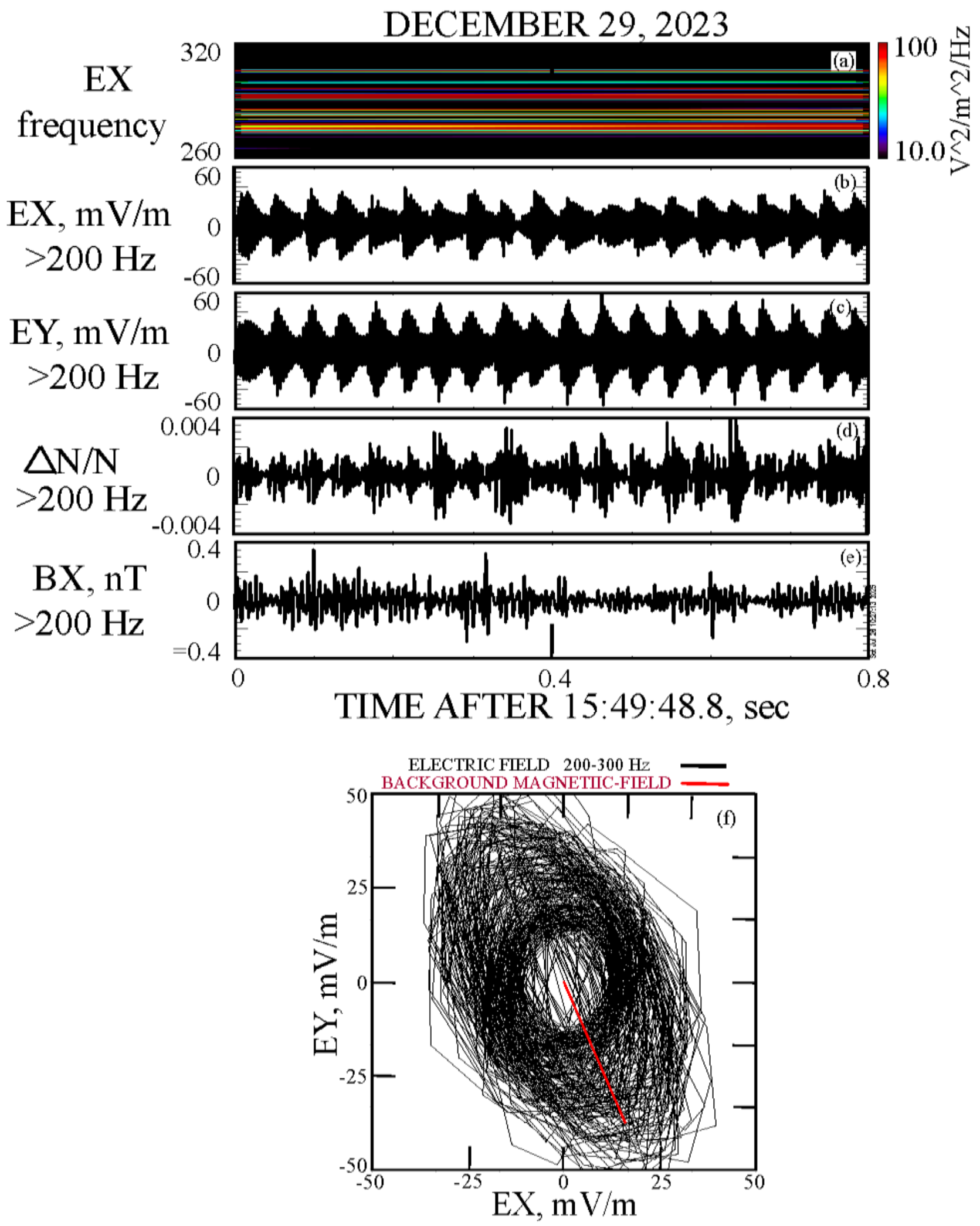

Figure 3. The high frequency wave of the TIAW is an electrostatic wave because it contains density fluctuations in panel 3d and no wave magnetic field in panel 3e. It propagates at a few degrees with respect to the background magnetic field (panel 3f). The high frequency electric field waveforms and spectra are illustrated in panels 3a, 3b, and 3c.

It may be important for the TIAW process being investigated that the low frequency wave is an electromagnetic wave, as shown in Figure 2i, because electromagnetic waves are known to play crucial roles in transferring energy and momentum in space and laboratory plasmas. Their compressible nature makes them highly effective at interacting with particles and other waves, leading to heating, acceleration, and the redistribution of energy across different scales [Porkolab and Bonoli, 1993; Yuan et al, 2018].

The existence of the TIAW depends on the ratio of the electron to ion temperature ($T_e/T_i$), as is illustrated in Figure 4, in which the higher frequency waves of panel (4a) and the density fluctuations associated with the low frequency wave in panel (4b) vary with the value of $T_e/T_i$ in panel (4c). Such may arise because ion acoustic waves are heavily damped at smaller values of $T_e/T_i$.

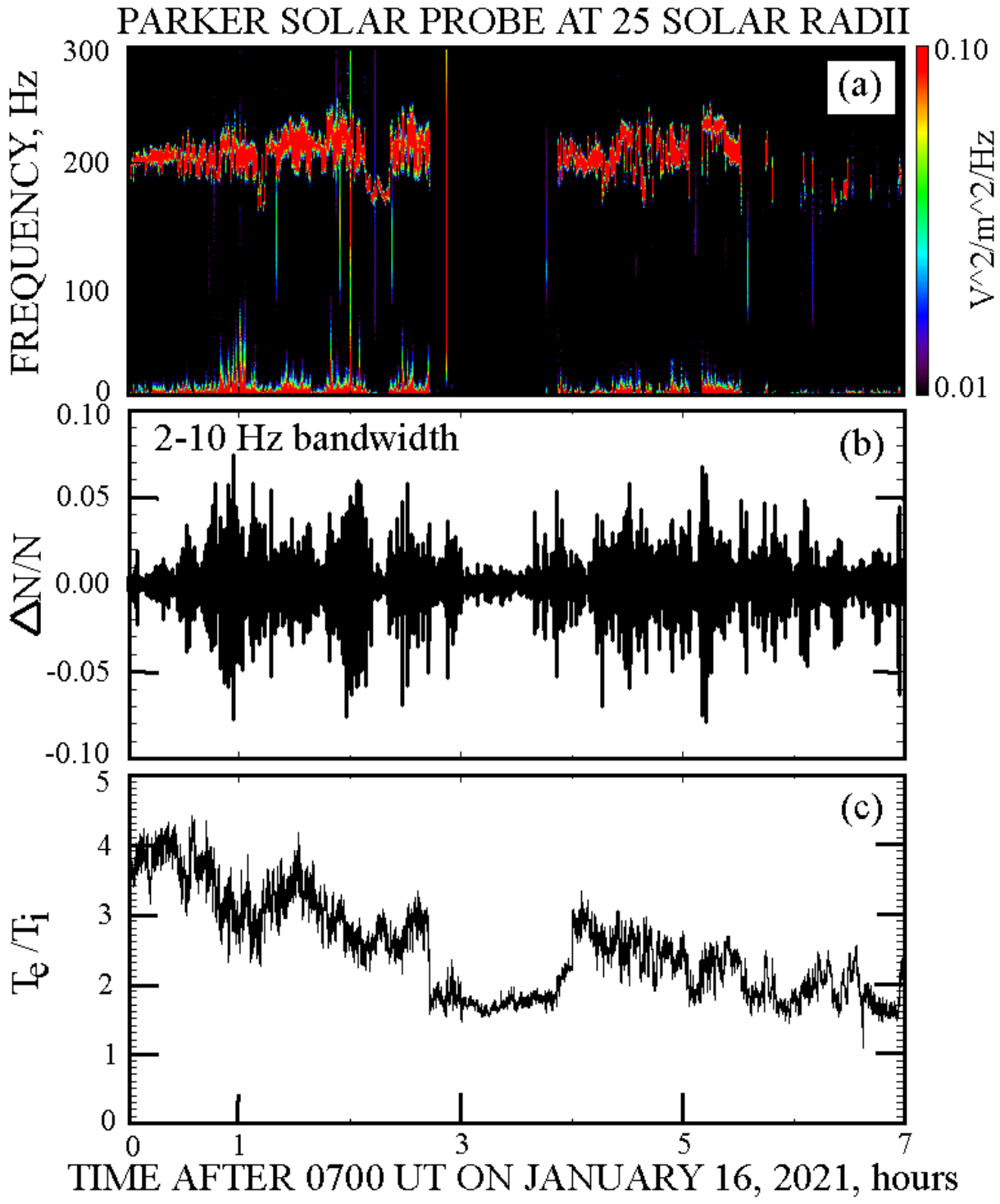

Figure 4. Panel (4a), the higher frequency wave spectra, Panel (4b) the higher frequency density fluctuation waveform and Panel (4c) the ratio of the electron to ion temperature. The breaks in the higher frequency spectrum and the minima in the density fluctuation amplitude occur when $T_e/T_i$ is at its minimum (from about 2.8 to 4 on the time scale) and maximum (from 0 to 0.5) values.

Figure 5 offers more information about the triggered ion acoustic waves. As seen in panels (5a) and (5b), both waves consist of harmonics whose frequency and amplitude depend on the plasma density of panel (5c). They all occur for intermediate values of $T_e/T_i$, shown in panel (5d).

Many TIAW events contain several higher frequency ion acoustic waves, such as that illustrated in Figure 6. The frequencies of the participating waves are given in the left part of the plots in this figure. The lowest frequency wave, at 4-10 Hz was the only wave having a magnetic field component so it was an electromagnetic wave and all the others were electrostatic waves. The vertical dashed lines through each peak of the highest frequency wave also pass through fixed phases of the lower frequency waves. This can only happen if the waves all had the same phase velocity.

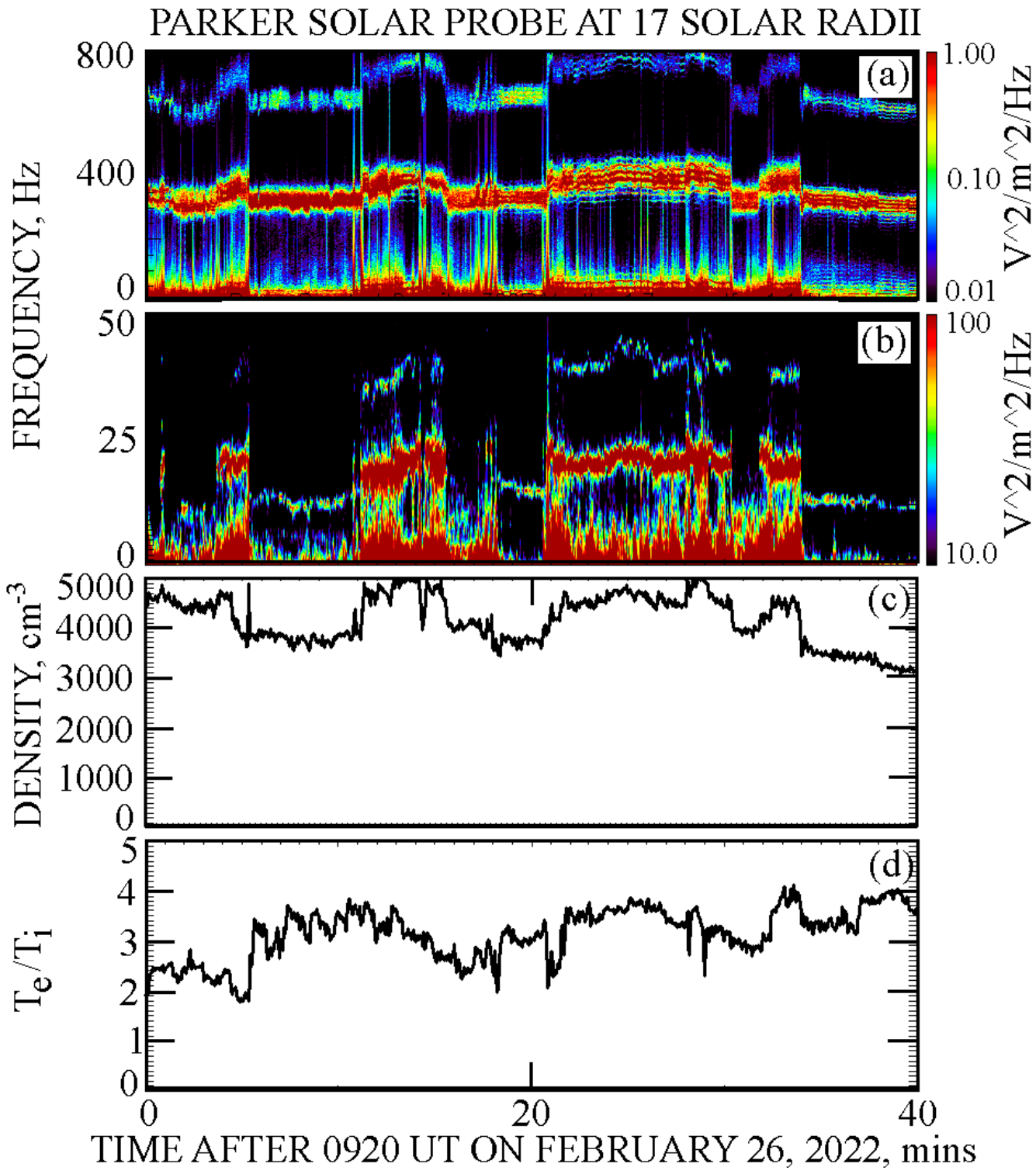

Figure 5. An interval during which both the high and low frequency waves of panels (5a) and (5b) are narrow band with many harmonics, and their frequencies and amplitudes are related to the plasma density of panel (5c). The ratio of the electron to ion temperatures is shown in panel (5d).

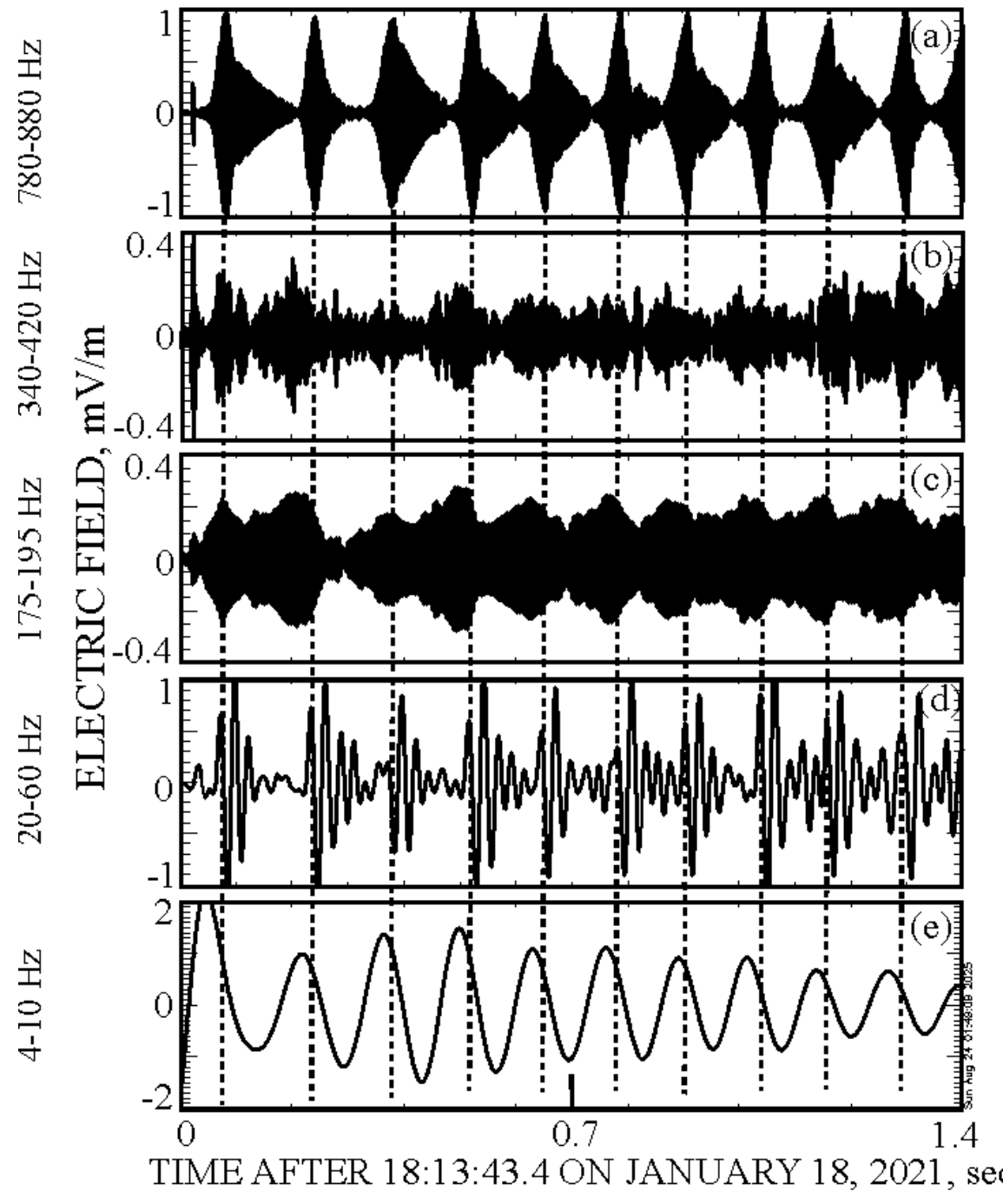

Figure 6. A triggered ion acoustic wave event with participating waves at the five frequencies illustrated at the left of the plots. The waveforms are illustrated in panels (a)

through (e). The dashed vertical lines show that the relative phases between all the waves were constant through the interval, which is strong evidence that all the waves have the same phase velocity.

This phase velocity may be estimated from the fact that the observed high frequency electrostatic wave, having a phase speed much less than the electron thermal speed, must have its parallel electric field balanced by the electron pressure gradient [Davidson, 1972; Kelley and Mozer, 1972], as in Equation (1).

$$eNE_{||} = -\nabla_{||} p_e \tag{1}$$

where N is the plasma density, $E_{||}$ is the parallel electric field (the component of the electric field parallel to the magnetic field direction), and $\nabla_{||} p_e$ is the parallel electron pressure gradient, $\nabla_{||} NkT_e$. Thus,

$$\delta\varphi / T_e \approx \delta N / N \tag{2}$$

$\delta\varphi$ depends on the value of the unmeasured EZ. For EZ to be unimportant in the estimation of $E_{||}$, the value of BZ should be small such that the magnetic field is in the X-Y plane. For the event described in Figure 7, BX = -20 nT, BY = 200 nT, and BZ < 1 nT. Panels (7a) and (7b) of Figure 7 present the higher and lower frequency TIAW waves during the interval in which BZ was small, as shown in panel (7c) $\Delta N/N$ was not measured for the event of interest. However, the two neighboring events had essentially identical parameters (except for the magnetic field direction), so their values of $\Delta N/N=0.0008$ are used in this analysis.

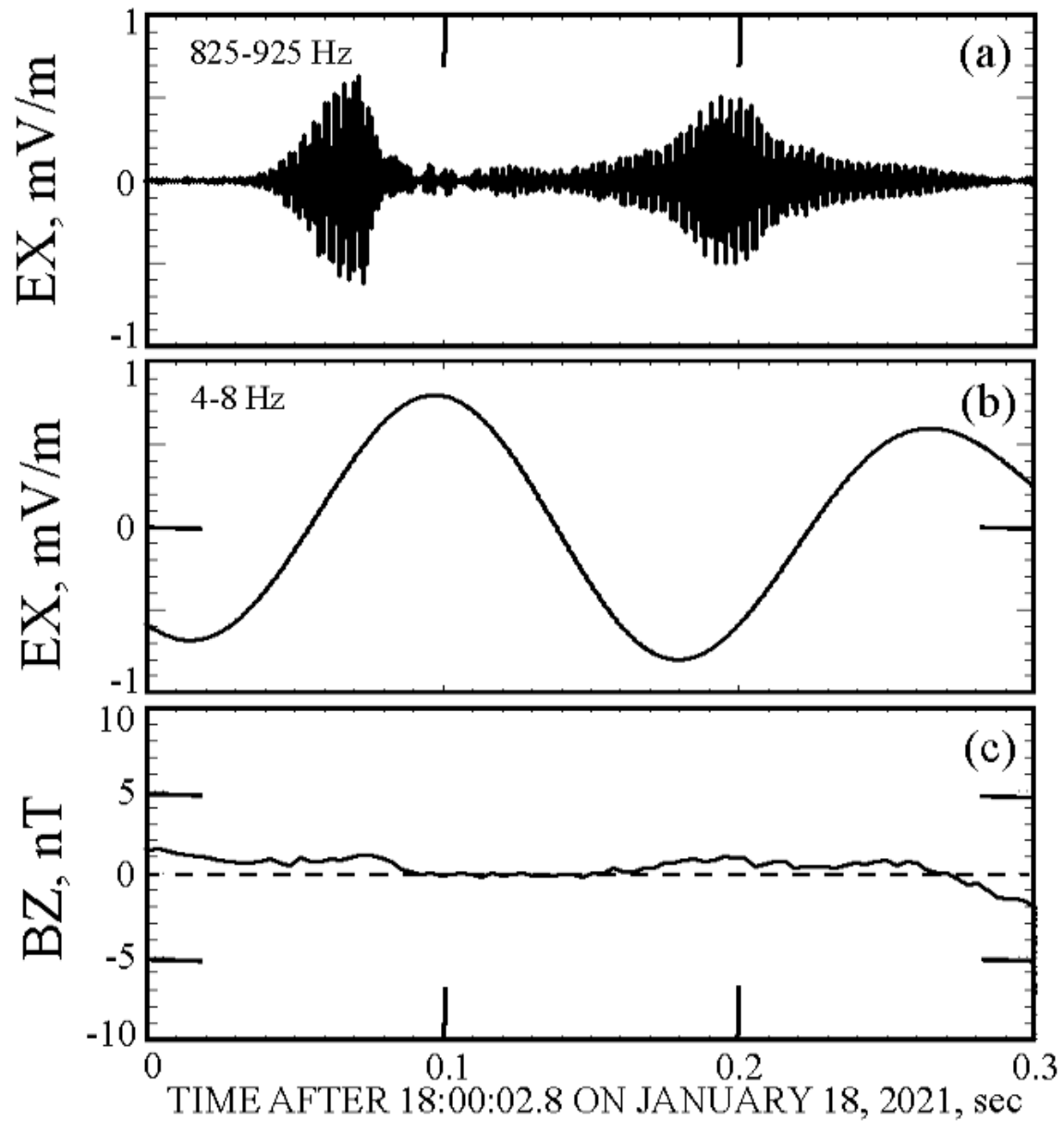

Figure 7. A TIAW event that occurred when the Z-component of the magnetic field was nearly zero, as seen in panel (7c). Panel (7a) gives the higher frequency electrostatic component of the electric field (filtered at 825-925 Hz) and panel (7b) gives the low frequency electromagnetic component as filtered from 4 to 8 Hz.

For EX =0.6 mV/m and EY = -0.6mV/m, the dot product of **E** and **B** gives the amplitude of the parallel electric field as

$$E_{||} = 0.6+EZ/200 \text{ mV/m}.$$

Neglecting the EZ term, the parallel potential, $\Delta\varphi$, may be estimated as

$$\Delta\varphi= < E_{||} >*(\text{length of half a wave period})$$

where $< E_{||} > = 2E_{||}/\pi$
and (length of half a wave period) = $(0.5/f)*v_{ph}$
where f = wave frequency = 875 Hz, and $v_{ph}$ is the wave phase velocity

With $T_e$ =40 eV, the only unknown in equation 2 is the wave phase velocity, which, from evaluating equation 2, is equal to 145 km /sec. During this time, the ion acoustic speed (cs) was about 100 km/sec.

While the observational link between the existence of the low frequency wave, its amplitude, the electron heating, and the presence of the high frequency TIAW is robust, the precise, step-by-step microphysical mechanism by which the effects of the low frequency wave specifically trigger the growth of the high frequency wave via a non-resonant interaction has not been understood. The following section describes an analysis that starts to produce this understanding.

## III PARTICLE-IN-CELL SIMULATION

In order to identify detailed processes leading to the generation of TIAW, we have performed a large number of full-kinetic (all species kinetic) simulations with varying parameters and the basic properties of the low-frequency (LF) wave. Here we describe the simplest example, which serves to illuminate the mechanism of the low-frequency wave triggering the growth of the high-frequency wave common to all cases. Since we use the model for qualitative illustration of the essential physics at play, we do not include quantitative comparisons. The parameters of the model (e.g. frequencies of the driving mode, amplitude of the electric field perturbations, etc.) were chosen to be within a reasonable range of the observations. The density variations selected are consistent (0.04-0.06) with the numerical results and the PSP observations.

The simulation demonstrates wave-particle interactions driven by the enhanced LF fluctuations, which subsequently modify the proton VDF and create conditions favorable for the growth of HF waves. Specifically, we consider an electrostatic LF wave propagating purely parallel to the background magnetic field. This model illustrates the same physics as in the case of an electromagnetic mode, namely that TIAWs are driven by modifications of the parallel distribution induced by the parallel electric field of a low frequency mode. By using purely parallel electrostatic mode, we could fit both LF and HF modes in the same 1D simulation. In contrast, using an electromagnetic LF mode would require much more computationally challenging 2D simulations. The simulation frame is collisionless, and quasi-neutral. The 1D simulation was performed using the fully kinetic particle-in-cell (PIC) code VPIC [Bowers, 2008]. The simulation domain has length $L_x \approx 29.3\, d_{ip}$ resolved with 7168 cells, where $d_{ip}$ is the ion inertial length. The initial distribution function was obtained from fitting the PSP observations and consists of three populations: electrons (e), core ions (c), and ion beam (b) [Mozer et al, 2021]. Each population is modeled by an isotropic Maxwellian drifting along the background magnetic field $\boldsymbol{B_0}$. The magnetic field is aligned with the simulation axis $z$. The ion beam density is $n_b = 0.11 n_e$ and core ion density is $n_c = 0.89 n_e$. The electron beta is $\beta_e = 8\pi n_e T_e / B_z^2 \approx 0.24$, $T_c/T_e \approx 0.35$, $T_b/T_e \approx 2.6$. The drifts are $V_c/v_{te} \approx -0.004$, $V_e = 0$, and $V_b = -V_c n_c/n_b$. Here $v_{te} = (T_e/m_e)^{1/2}$. The ratio of the electron plasma frequency to the electron cyclotron frequency is $\omega_{pe}/\Omega_{ce} = 20$, ion-to-electron mass ratio is $m_i/m_e = 900$, and time step is $\delta t = 0.12\omega_{pe}$. The number of particles per cell is very large, $1.6 \times 10^5$ per species, to suppress the numerical noise. An electrostatic LF wave with frequency $\omega_0 = 0.94\Omega_{ci}$ and wavenumber $k_0 = \omega_0/C_s$, where $C_s = (T_e + 3T_c)^{1/2}/m_i^{1/2}$ is driven by an externally imposed current system. The amplitude of the electrostatic LF wave was tuned to keep wave perturbations of plasma density on the level of a few percent of the background solar wind plasma density as seen in the PSP observations. Figures 8a-d shows the snapshot of the simulation.

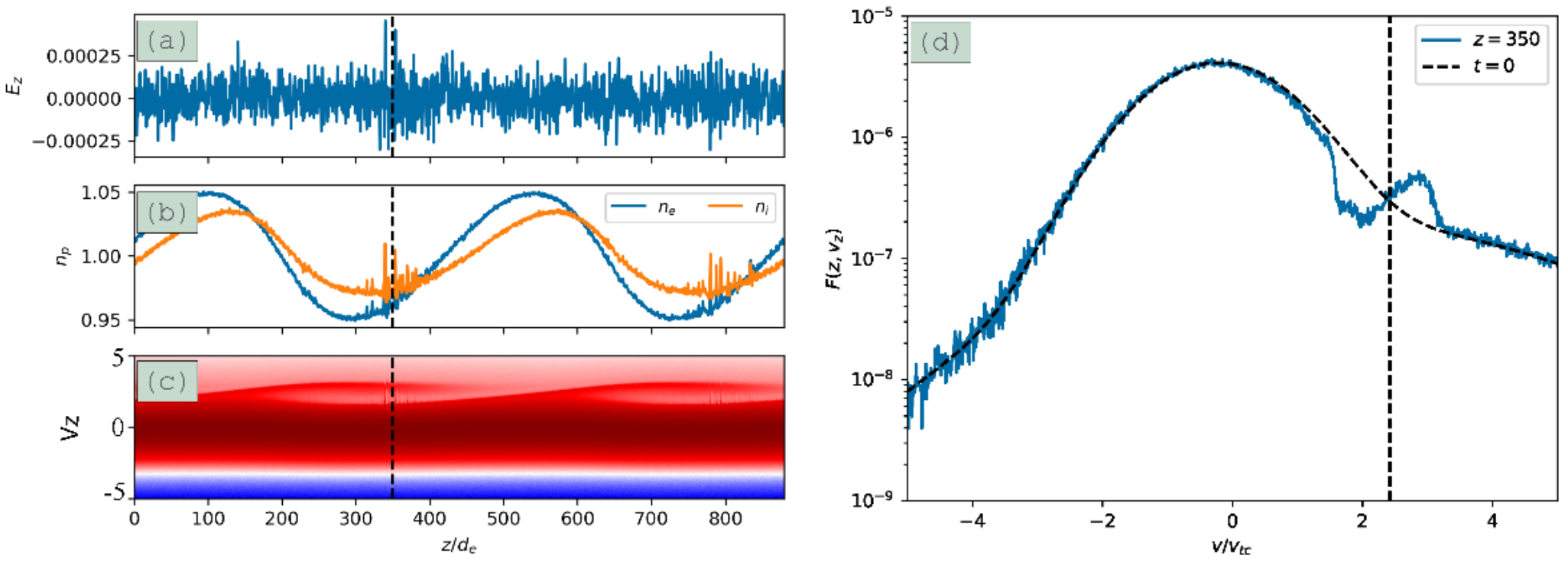

Figure 8. Results of 1D PIC simulation. (a) – the parallel electric field perturbations at time $t\omega_0 = 26$; (b) – the proton density normalized to the reference density ; (c) – the distribution function of protons in the $z/d_e$ and $v_z$ domain. (d) - the proton distribution function at location= $350\, d_e$: initial – the dashed curve, current – the blue curve. Here $d_e = c/\omega_{pe}$ is the electron inertial length and $c$ is the speed of light.

The electrostatic component of the LF wave, in Figure 8a, acts as a slowly varying background potential and reshapes the proton distribution function $f_p(z, v_z, t)$ as seen in Figures 8c, and 8d. Also note that the proton and electron densities of Figure 8b develop perturbations at a fixed phase of the wave, which is like the situation observed for the high frequency wave of the TIAW. Steepening of the electric field and plasma density LF perturbations is also observed, as often seen in the PSP electric field and plasma density data (Figures 1 and 2). That is similar to the modulation of the electron distribution function by whistler waves discussed in [Drake et al. 2015; Agapitov et al. 2018]. This modulation effectively redistributes proton phase-space density in regions where the LF fields are strongest. Regions of enhanced parallel electric field associated with the LF mode can modify parallel ion velocity distribution, creating localized plateaus or crests in $f_p$. In particular, the LF modulation produces spatial regions where $f_p$ has beam-like features around the ion acoustic phase speed that are favorable for growth of higher-frequency (HF) ion acoustic electrostatic waves. A key consequence of this is the generation of HF ion acoustic-like electrostatic waves, with the amplitudes modulated by the LF wave phase. That process is sensitive to the electron-to-ion temperature ratio $T_e/T_i$, since generation of beam-driven electrostatic mode in the ion-acoustic family requires a relatively large value of $T_e/T_i$. If $T_e/T_i$ is small, the resonance moves deeper into the core, and the damping of the LF wave is too strong for it to exist. If $T_e/T_i$ is too large, the resonance moves too far into the tail of the distribution.

In the nonlinear regime, the beam driven ion acoustic mode inherits a mixed character: it bears signatures of both classic ion acoustic waves and nonlinear perturbations like ion acoustic solitons both tending to flatten the local positive slope on $f_p$ similarly to the electron scale time domain structures in the Earth's radiation belts, nonlinear electron acoustic mode [Mozer et al. 2015, 2016; Vasko et al. 2017a, b]. The proton distribution function thus participates actively in the nonlinear energy exchange: the LF electrostatic wave reshapes $f_p$, which modulates and promotes locally the growth of HF ion acoustic structures; these, in turn, feedback heating the protons, flattering the $f_p$, and slowing down steepening of the LF field.

## IV SUMMARY

Triggered ion acoustic waves (TIAWs) are the dominant emissions observed below 10,000 Hz at 10-30 solar radii. They are characterized by the simultaneous presence of two distinct coupled components: a low-frequency (LF) wave typically oscillating at a few Hz, and high-frequency (HF) wave with frequencies of a few hundred Hz. All the waves in a TIAW event travel at the same phase speed, estimated to be 150 km/sec. Their existence and coupling depend on $T_e/T_i$, the ratio of the electron to ion temperature. Particle-in-cell (PIC) models were used to illustrate the interplay between the pair of waves mediated by modifications of plasma parameters and nonlinear energy conversion.

The understanding that low frequency waves can drive the growth of higher frequency waves in a non-resonant process has several important implications and opens avenues for future investigation:

● Plasma Heating and Energy Transport: The TIAW phenomenon highlights a significant and potentially widespread mechanism for ion and electron heating in space plasmas like the solar wind. This has direct consequences for models of solar wind acceleration, thermal evolution, and energy partitioning.

● Microphysics of Wave Coupling: While the TIAW observations establish the link between low frequency waves, electron heating, and high frequency wave generation, the detailed microphysics of this coupling—specifically, how the low frequency wave-modified plasma state leads to the precise characteristics of the high frequency wave bursts—remains to be fully explored. Further theoretical modeling and advanced numerical simulations are needed to pinpoint the exact coupling mechanisms and to quantitatively reproduce the observed wave characteristics.

## V ACKNOWLEDGEMENTS

This work was supported by NASA contract NNN06AA01C. The authors acknowledge the extraordinary contributions of the Parker Solar Probe spacecraft engineering team at the Applied Physics Laboratory at Johns Hopkins University. The FIELDS experiment on the Parker Solar Probe was designed and developed under NASA contract NNN06AA01C. VR was supported by NASA grant 80NSSC22K1014. OVA was supported by NASA contracts 80NSSC22K0433, 80NNSC19K0848, 80NSSC21K1770, and NASA's Living with a Star (LWS) program (contract 80NSSC20K0218). This research used resources provided by the NASA HighEnd Computing (HEC) Program through the NASA Advanced Supercomputing (NAS) Division at Ames Research Center.

## REFERENCES

Agapitov O., Drake J. F., Vasko I., Mozer F. S., Artemyev A., Krasnoselskikh V., et al. (2018). Nonlinear Electrostatic Steepening of Whistler Waves: The Guiding Factors and Dynamics in Inhomogeneous Systems. *Geophysical Research Letters*, *0*(0). https://doi.org/10.1002/2017GL076957

Bale, S.D., Goetz, K., Harvey, P.R., Turin, P. Bonnell, J. W., Dudok de Wit, T., Ergun, R.E., MacDowall, R. J., Pulupa, M., et al, (2016) "The Fields Instrument Suite for Solar Probe Plus", SSRv, 204, 49

Bowers, K. J., Albright, B. J., Yin, L., Bergen, B., & Kwan, T. J. T. (2008). Ultrahigh performance three-dimensional electromagnetic relativistic kinetic plasma simulation. *Physics of Plasmas*, *15*(5), 055703. DOI: https://doi.org/10.1063/1.2840133

Cattell, C., Breneman, A., Dombeck, J., et al (2021). Parker Solar Probe evidence for the absence of whistlers close to the Sun to scatter strahl and regulate heat flux. *The Astrophysical Journal Letters*, *911*(1), L29.

Davidson, R.C., (1972), Methods in nonlinear plasma theory, eBook ISBN: 9780323153386, Elsevier)

Drake, J., Agapitov, O., & Mozer, F. (2015). The development of a bursty precipitation front with intense localized parallel electric fields driven by whistler waves. *Geophysical Research Letters*, *42*(8), 2563–2570.

Jian, L.K., Wei, H.Y., Russell, C.T., J. G. Luhmann, J.G. Klecker, B., et. Al, (2014), Electromagnetic waves near the proton cyclotron frequency: STEREO observations, The Astrophysical Journal, 786:123, doi:10.1088/0004-637X/786/2/123

Kelley, M. C. and Mozer, F. S., (1972), A technique for making dispersion relation measurements of electrostatic waves, Journal of Geophysical Research, vol. 77, no. 34, pp. 6900–6903. doi:10.1029/JA077i034p06900.

Mozer, F. S., Bale, S. D., Cattell, C. A., et al, (2021). Triggered ion acoustic waves in the solar wind. ApJL,,919 L2.

Mozer, F. S., Bale, S. D., Cattell, C. A., et al (2022). Core Electron Heating By Triggered Ion Acoustic Waves, The Astrophysical Journal Letters, 927:L15, https://doi.org/10.3847/2041-8213/ac5520

Mozer, F.S., Bale, S.D., Kellogg, P.J. et al, (2023) Arguments for the physical nature of the triggered ion-acoustic waves observed on the Parker Solar Probe, Phys. Plasmas 30, 062111 doi: 10.1063/5.0151423

Mozer, F.S., Agapitov, O.V., Choi, K-E., et al, (2025) Parallel Electric Fields and Electron Heating Observed in the Young Solar Wind, The Astrophysical Journal, 981:82 https://doi.org/10.3847/1538-4357/adb582

Ofman, L., Nakariakov, V. M. & Deforest, C. E. (1999) Slow magnetosonic waves in coronal plumes. Astrophys. J. 514, 441–447. doi:10.1086/306944

Ofman, L., Wang, T.J., and Davila, J.M., (2012), Slow magnetoionic waves and fast flows in active region loops, The Astrophysical Journal, 754:111, doi:10.1088/0004-637X/754/2/111

Porkolab, M., & Bonoli, P. T. (1993). *Plasma Heating by Fast Magnetosonic Waves in Tokamaks*. MIT Plasma Science and Fusion Center, Report PFC/JA-93-19. Handle: http://hdl.handle.net/1721.1/95179

Vasko, I. Y., Agapitov, O. V., Mozer, F. S., Artemyev, A. V., Drake, J. F., & Kuzichev, I. V. (2017). Electron holes in the outer radiation belt: Characteristics and their role in electron energization. *Journal of Geophysical Research: Space Physics*, *122*(1), 2016JA023083. https://doi.org/10.1002/2016JA023083

Vasko, I. Y., Agapitov, O. V., Mozer, F. S., Bonnell, J. W., Artemyev, A. V., Krasnoselskikh, V. V., et al. (2017). Electron-acoustic solitons and double layers in the inner magnetosphere. *Geophysical Research Letters*, *44*(10), 2017GL074026. https://doi.org/10.1002/2017GL074026

Whittlesey, P.L., Larson, D.E., Kasper, J.C. Halekas, J., Abatcha, M., Abiad, R., Berthomier, M., Case, A.W., Chen, J., Curtis, D.W., et al, (2020) The Solar Probe ANalyzers—Electrons on the Parker Solar Probe, The Astrophysical Journal Supplement Series, 246:74 (14pp), February https://doi.org/10.3847/1538-4365/ab7370

Yuan, Z., Yu, X., Huang, S., Qiao, Z., Yao, F., & Funsten, H. O., (2018), Cold Ion Heating by Magnetosonic Waves in a Density Cavity of the Plasmasphere, Journal of Geophysical Research: Space Physics, 123, Issue 2, https://doi.org/10.1002/2017JA024919

Zieger, B., Zank, G. P., Adhikari, L., & Florinski, V. (2020) Dispersive Fast Magnetosonic Waves and Shock-Driven Compressible Turbulence in the Inner Heliosheath, *Journal of Geophysical Research: Space Physics*, 125, 12, **DOI:** https://doi.org/10.1029/2020JA028393